# Classifying chemical elements and particles: from the atomic to the sub-atomic world

Maurice R Kibler

Institut de Physique Nucléaire
IN2P3-CNRS et Université Claude Bernard Lyon-1
43, Boulevard du 11 Novembre 1918
69622 Villeurbanne Cedex, France

**Abstract**
This paper presents two facets. First, we show that the periodic table of chemical elements can be described, understood and modified (as far as its format is concerned) on the basis of group theory and more specifically by using the group SO(4,2)xSU(2). Second, we show that "periodic tables" also exist in the sub-atomic and sub-nuclear worlds and that group theory is of paramount importance for these tables. In that sense, this paper may be considered as an excursion, for non specialists, into nuclear and particle physics.

**In memory of Jean Gréa**
This paper is dedicated to the memory of Jean Gréa who, like Harry Wiener, was a universal man. Jean was a physicist with a large flow of activities in theoretical physics. He started with nuclear physics, then switched to pre-quantum mechanics (theory of hidden variables), continued with simulations in physics and the theory of signal and finally made important contributions in didactics and philosophy of sciences. He was the co-founder of the LIRDHIST (*Laboratoire Interdisciplinaire de Recherche en Didactique et en Histoire des Sciences et Techniques*) at the *Université Claude Bernard Lyon 1*. As a teacher, he knew how to communicate his enthusiasm to students. He was very much oriented to other people. He knew how to listen to others, how to answer questions by sometimes asking other questions and how to help people. *Merci Jean.*

# Classifying chemical elements and particles: from the atomic to the sub-atomic world

Maurice R Kibler

Institut de Physique Nucléaire
IN2P3-CNRS et Université Claude Bernard Lyon-1
43, Boulevard du 11 Novembre 1918
69622 Villeurbanne Cedex, France

## 1 Introduction

Holding a conference in 2003 on the periodic table of chemical elements may seem somewhat curious at first sight. The periodic table has been the object of so many studies that it is hard to imagine any new developments on the subject except perhaps from the perspective of History, Scientific Philosophy, Epistemology, Sociology or Politics.

From the point of view of Chemistry and Physics, what can be said about the periodic table in the 21st century? Along this vein, we may ask the following questions.

- Does the periodic table have a limit?
- Do other formats in two or three dimensions provide insight for teaching purposes?
- What can be learned from periodic tables in other dimensions (as for example in Flatland, a two-dimensional space, or in a four-dimensional space)?
- What can be gained from a quantum-mechanical approach and what is the importance of relativistic quantum mechanics for the periodic table?
- What can be gained from a group theoretical approach to the periodic table?
- What are the implications of a group theoretical approach for molecules?
- How may the periodic table of neutral atoms be extended to a periodic table for ions or to a periodic table for molecules?
- What are the analogues of the periodic table in the sub-atomic world and in the sub-nuclear world?

Many of these questions were addressed during the Second Harry Wiener International Memorial Conference and the reader will find some answers in this volume and in the other publications connected to the conference.

This paper deals with two of the preceding questions. First, we show that the periodic table of chemical elements can be described, understood and modified (as far as its format is concerned) on the basis of group theory and more specifically by using the group SO(4,2)xSU(2). Second, we show that “periodic tables” also exist in the sub-atomic and sub-nuclear worlds and that group theory is of paramount importance for these tables. In that sense, this paper may be considered as an excursion into nuclear and particle physics.

Although group theory constitutes an important tool for the classification of chemical elements and particles, no special knowledge of this branch of mathematics is required for what follows. The presentation is self-contained and emphasizes numerous historical facets. In addition, the so-called *eka-process* (i.e., the prediction of a new element or a new particle with some of its properties with *a posteriori* verifications) constitutes a leitmotiv of this work.

The paper is organized in the following way. Section 2 is devoted to the notion of elements from Antiquity to the year 2003. In Section 3, some basics about group theory are presented with the types of groups of interest not only for the periodic table but also for the standard model of particles and their interactions. Section 4 is concerned with an SO(4,2)xSU(2) approach to the periodic table of neutral elements. The transition to fundamental particles is made in Section 5. Finally, some concluding remarks are presented in Section 6.

## 2 Elements from Antiquity to 2003

### *2.1 From Antiquity to the 18th century*

The idea of first principles (elements) as primary constituents of matter is not new. In this respect, the Greek philosophers of Antiquity are a good starting point without denying that similar ideas could have been developed at other times by other people. For the ancient Greeks, any creation or annihilation proceeded *via* one or more first principles.

The notion of a single element as the basis of everything was advanced by Thales (water), by Anaximenes and Diogenes (air) and by Hippasos and Heraclitus (fire). Later, Empedocles considered four elements (fire, earth, water, air) and was original not only in increasing the number of elements from one to four but also in introducing two mediating objects (interactions)[1] between the elements of the quartet. These mediating objects, love (attraction) and hate (repulsion) could differentiate and combine the elements. In Plato's treatment, the four elements were put into one-to-one correspondence with four of the regular polyhedrons: fire $\leftrightarrow$ tetrahedron, earth $\leftrightarrow$ cube, water $\leftrightarrow$ icosahedron and air $\leftrightarrow$ octahedron. (The fifth regular polyhedron - the dodecahedron - was supposed to correspond to the whole Universe.) Finally, Aristotle introduced the quintessence (aether) as a fifth element. According to Leucippus and his student Democritus, the world consisted of atoms with two basic elements (full or "to be" and void or "not to be") while according to the Pythagorean school, the origin of everything is to be found in mathematical principles.

The idea of four or five elements was still accepted, with minor modifications, up to the Middle Ages and, to a certain extent, into the fifteenth and sixteenth centuries (*la Renaissance*), the seventeenth century (*le Grand Siècle*) and part of the eighteenth century (*le Siècle des Lumières*). Apart from alchemy and the phlogiston theory of fire, no new ideas about the constitution of matter appeared for centuries.

[1] Note the parallel between the notion of mediating objects and the concept, in modern theoretical physics, of interaction or gauge fields (the quanta of which are the so-called intermediate or gauge bosons) between particles.

### *2.2 From Lavoisier to Mendeleev*

The situation slowly changed with the introduction of experiments. Lavoisier abandoned the idea of phlogiston in 1783 and in his celebrated *Traité Élémentaire de Chimie* in 1789 gave a list of 33 simple substances containing 23 chemical elements and some other substances including "light" and "caloric". The discovery of new elements continued as shown in the following table giving the number of elements known at a given year and the name of a scientist representative of the year.

| number of elements | year | representative scientist |
|---|---|---|
| 23 | 1789 | Lavoisier |
| 30 | 1815 | Prout |
| 40 | 1818 | Berzelius |
| 49 | 1828 | Berzelius |
| 61 | 1849 | Gmelin |
| 63 | 1865 | Meyer, Mendeleev |

As a result of the increasing number of chemical elements, many scientists in the middle of the nineteenth century tried to develop classification schemes based on similarities, trends and periodicity of chemical properties. The early attempts include Döbereiner's triads (1817) and their expansion by Gmelin as series (1827), Pettenkofer's groupings (1850), Beguyer de Chancourtois' spiral or *vis tellurique* (1862), Newlands' octaves (1865) and the tables by Olding (1865) and Meyer (1864-70). Döbereiner's triads

(Cl, Br, I) – (S, Se, Te) – (Ca, Sr, Ba) – (Li, Na, K) – (Fe, Co, Ni)

and the spiral graph of Beguyer de Chancourtois already contained some ingredients and trends of the periodic table. Beguyer de Chancourtois arranged the elements in increasing atomic weight on a spiral drawn on a right cylinder with a periodicity of order 8. All the elements exhibiting similar properties were located on the same generatrix of the spiral. Newlands' classification into the octaves

| | | | | | | |
|---|---|---|---|---|---|---|
| H | Li | Be | B | C | N | O |
| F | Na | Mg | Al | Si | P | S |
| Cl | K | Ca | Cr | Ti | Mn | Fe |

in 1864 also emphasizes the notion of periodicity of chemical and physical properties of elements.

A few years after the work of Beguyer de Chancourtois and Newlands, Meyer and Mendeleev gave a more elaborate form to the classification. More specifically, Mendeleev set up in his "*Essai d'une (sic) Système des Eléments*" in 1869 a periodic system of elements in groups and rows, where elements are disposed in order of increasing atomic weight and chemically similar elements are placed beneath one another. Such a system comprised ten horizontal rows (corresponding to seven periods) and eight vertical columns (corresponding to eight, rather than nine, groups because at that time the group of rare gases was not known).

The periodic table by Mendeleev was based on the periodicity of chemical properties of the 63 elements (out of the 92 natural elements) known at that time. As a matter of fact, in order to preserve the periodicity of their chemical properties, Mendeleev found it necessary to transpose some elements against the order of increasing atomic weight, e.g., using the notation “symbol(Z=atomic number, atomic weight)”, I(Z=53, 126.9) $\leftrightarrow$ Te(Z=52, 127.6) and Ni(Z=28, 58.7) $\leftrightarrow$ Co(Z=27, 58.9). Mendeleev’s work was effective not only for simple classification and rationalization purposes but also for prediction purposes. Indeed, in his 1869 table, in order to maintain the periodicity of the chemical properties, Mendeleev left empty boxes and predicted the existence of some new elements, the so-called eka-elements, together with their main properties, approximate atomic mass and production process. In particular, in a column between Si(Z=14) and Sn(Z=50), he predicted the occurrence of an element (which he called eka-Silicon) with an approximate atomic mass of 72. Such an element was isolated in 1886 by Winkler and is now called Germanium [Ge(Z=32)]. He did similar predictions in introducing eka-Aluminum and eka-Boron corresponding to the elements Gallium [Ga(Z=31)] and Scandium [Sc(Z=21)] isolated in 1875 and 1879, respectively. The next table illustrates the *eka-process* for Ga and Ge.

| | | | | | | |
|---|---|---|---|---|---|---|
| H | | | | | | |
| Li | Be | B | C | N | O | F |
| Na | Mg | Al | Si | P | S | Cl |
| K | Ca | empty box → prediction of eka-Aluminum | empty box → prediction of eka-Silicon | As | Se | Br |

The predictions of Ga and Ge were successful not only from a qualitative point of view (prediction of new elements) but also from a quantitative point of view (prediction of properties and data).

It is clear that the periodic table cannot be credited to a single author. In fact, Mendeleev was not the sole person to use atomic weights[2] as well as similarities, trends and periodicity for classifying chemical elements. In this regard, Beguyer de Chancourtois, Newlands and others deserve mention. However, Mendeleev stands out for predicting the existence of eka-elements qualitatively and some of their properties quantitatively. (The reader should consult the paper by D. Rouvray in this volume for further details.)

[2] A basic ingredient inherent to the periodic table by Mendeleev is the classification according to increasing atomic weight. It was recognized at the beginning of the twentieth century that the central rôle is played by the atomic number Z rather than by the atomic weight.

## *2.3 From 1870 to 2003*

Since Mendeleev's prediction in 1869 of new chemical elements (eka-Aluminum, eka-Silicon and eka-Boron), the size of the periodic table has increased as shown in the following table.

| year | 1870 | 1940 | 1958 | 1973 | 1987 | 2003 |
|---|---|---|---|---|---|---|
| number of known chemical elements | 70 | 86 | 102 | 105 | 109 | 114 |

The year 1958 indicates the limit of classical chemistry with the production of 50000 atoms of Nobelium (Z = 102, lifetime of a few minutes) and its doubly ionized cation in solution. For Z ≥ 93, the focus has shifted away from ordinary chemistry and the bonding between atoms, ions & molecules, and moved towards nuclear chemistry. So far 114 elements up to Z = 116 have been established; elements Z = 113 and Z = 115 are missing. The tables below give the year of discovery and other details (atomic number Z and symbol) for each isolated transuranic element from Z = 93 to Z = 116.

| Z | 93 | 94 | 95 | 96 | 97 | 98 |
|---|---|---|---|---|---|---|
| element | Neptunium | Plutonium | Americium | Curium | Berkelium | Californium |
| symbol | Np | Pu | Am | Cm | Bk | Cf |
| year | 1940 | 1940 | 1944 | 1944 | 1949 | 1950 |

| Z | 99 | 100 | 101 | 102 | 103 | 104 |
|---|---|---|---|---|---|---|
| element | Einsteinium | Fermium | Mendeleevium | Nobelium | Lawrencium | Rutherfordium |
| symbol | Es | Fm | Md | No | Lr | Rf |
| year | 1952 | 1952 | 1955 | 1958 | 1961 | 1973 |

| Z | 105 | 106 | 107 | 108 | 109 | 110 |
|---|---|---|---|---|---|---|
| element | Dubnium | Seaborgium | Bohrium | Hassium | Meitnerium | not named |
| symbol | Db | Sg | Bh | Hs | Mt | / |
| year | 1968 | 1974 | 1981 | 1984 | 1982 | 1994 |

| Z | 111 | 112 | 113 | 114 | 115 | 116 |
|---|---|---|---|---|---|---|
| element | not named | not named | not observed | not named | not observed | not named |
| symbol | / | / | / | | / | / |
| year | 1994 | 1996 | / | 1999 | / | 1999 |

## *2.4 Comparison with particle physics*

The present situation in chemistry may be compared with that in particle physics. Chemistry, including nuclear chemistry and bio-chemistry, depends upon 114 elements and embraces all known substances whatever their composition and structure (gases, liquids, crystalline and amorphous solids). The current situation in particle physics appears somewhat simpler. According to the current model of particles and their interactions, all matter including that produced in accelerators depends upon only 12 elementary matter particles + 13 gauge particles + 1 "feeding particle" plus the associated anti-particles.

The matter particles consist of six quarks and six leptons. The quarks are the constituents of the hadrons which comprise baryons (protons, neutrons, etc.) and mesons (pions, kaons, etc.). In this respect, we know the familiar descent

| |
|---|
| molecule: $10^{-7}$ cm $\rightarrow$ atom: $10^{-8}$ cm $\rightarrow$ nucleus: $10^{-12}$ cm $\rightarrow$ nucleon (proton or neutron): $10^{-13}$ cm $\rightarrow$ quark: $10^{-18}$ cm $\rightarrow$ ? |

according to decreasing dimension scales. The six leptons can be divided into three charged leptons (one of them is the electron) and three neutral leptons (one of them is the neutrino associated with the electron). Both quarks and leptons are fermions, obeying the Fermi-Dirac statistics, of spin ½.

The quarks and leptons interact *via* four interactions: (i) the strong interaction which ensures the cohesion of quarks inside the proton and the neutron and, in a residual form, the cohesion of protons and neutrons inside the nucleus; (ii) the electromagnetic interaction between charged particles; (iii) the weak interaction responsible for β-decay; (iv) the gravitational interaction between massive particles. The interactions between matter particles are made possible by thirteen "gauge" fields[3] with which are associated particles (8 gluons, 1 photon, 3 intermediate bosons and 1 graviton). Not all matter particles are responsive to the four interactions as shown in the following table.[4]

| interaction | relative strength | range (cm) | quanta of the field | concern | typical manifestation |
|---|---|---|---|---|---|
| strong | 1 | $10^{-13}$ | 8 gluons | quarks | nuclear forces |
| electro-magnetic | $10^{-2}$ | $\infty$ | photon | charged particles | atomic and molecular forces |
| weak | $10^{-4}$ | $10^{-15}$ | $W^-$, $W^+$, $Z^0$ | quarks & leptons | radioactivity (β-decay) |
| gravitational | $10^{-39}$ | $\infty$ | graviton (not observed) | massive particles | terrestrial and celestial mechanics |

Eight gluons are used to describe the strong interaction and three intermediate bosons ($W^-$, $W^+$ and $Z^0$) the weak interaction while the photon and the graviton serve for describing the electromagnetic interaction and the gravitational interaction, respectively. The mediating particles are bosons, obeying the Bose-Einstein statistics, of spin 1 (for the gluons, intermediate bosons and photon) or spin 2 (for the graviton). It is recognized that the gravitational interaction is negligible with respect to the three others at the atomic and sub-atomic scales. Therefore the gravitational interaction, responsible for the stability of the Universe, will not be discussed any longer in this paper.

---

[3] Think of the electromagnetic field and the gravitational field as examples.

[4] The neutrinos do not undergo strong and electromagnetic interactions. The electron-neutrino flux on earth is $10^{11}$ neutrinos/cm$^2$/s so that $10^{14}$ neutrinos go through the human body every second but only one neutrino is stopped by the body during its human life.

Finally, a last field needs to be introduced, namely, the Higgs field which is responsible for the mechanism providing mass to particles. Indeed, according to the standard model of particles and interactions in its former form, all particles have a zero mass. The Higgs mechanism makes it possible for certain particles (like the electrons, protons and neutrons) to acquire a mass. With the Higgs field is associated the Higgs boson which is a scalar particle of spin 0 (a "feeding" particle sometimes referred to as the God particle). The latter particle has not been observed and is still the object of experimental investigations.

## 3 Group theory in Chemistry and Physics

### *3.1 Groups*

The mathematical structure of a group is certainly one of the simplest algebraic structures useful in Chemistry and Physics. A group is a set of elements endowed with an internal composition rule (i.e., the combination of two arbitrary elements in the set produces a unique element of the set) which is associative, has a neutral element and is such that each element in the set has an inverse with respect to the neutral element. As a trivial example, the set of symmetries in the 3-dimensional space $\mathbf{R}^3$ that leave a regular octahedron invariant is a group known in Chemistry as O (if only rotations are involved) or $O_h$ (if rotations and the inversion are involved). As another example, the set of rotations in $\mathbf{R}^3$ that leave invariant the real form $x^2+y^2+z^2$ is a group noted SO(3).[5] Similarly, the space–time symmetries in the 3+1-dimensional space $\mathbf{R}^{3,1}$ (restricted to rotations in $\mathbf{R}^3$ and Lorentz transformations in $\mathbf{R}^{3,1}$) leaving the real form $x^2+y^2+z^2-c^2t^2$ is a group noted SO(3,1).

These examples clearly show that it is generally possible to endow a set of symmetries with a group structure by using the composition of symmetries as an internal law. According to the cardinality of the set, we may have

- discrete groups (having a finite number or an infinitely countable number of elements), like the 32 crystallographic point groups of molecular physics or the 230 space groups of solid state physics, and
- continuous groups (having an infinite number of elements), like the rotation group SO(3) and the Lorentz group SO(3,1).

Among the continuous groups, we have to distinguish the compact groups and the noncompact groups. Each element of a continuous group can be characterized by a finite number of parameters. (For example, we need three angles for characterizing the rotation of a given angle around a given axis.) If each parameter varies on a compact (i.e., closed and bounded) interval, the group is said to be compact. It is noncompact in the opposite case. As examples of compact groups, we have

- SO(n) the special orthogonal group in n dimensions (the elements of which are orthogonal real nxn matrices of determinant +1)

---

[5] The notation comes from the fact that this group is isomorphic with the set of orthogonal real 3x3 matrices of determinant unity endowed with matrix multiplication.

- SU(n) the special unitary group in n dimensions (the elements of which are unitary nxn matrices of determinant +1).

Among the noncompact groups, we may quote

- SO(3,1) the Lorentz group in 3+1 dimensions
- SO(3,2) the de Sitter groups in 3+2 dimensions
- SO(4,1) the de Sitter group in 4+1 dimensions
- SO(4,2) the conformal group in 4+2 dimensions

(the elements of SO(p,q) are pseudo-orthogonal real (p+q)x(p+q) matrices of determinant +1 with the metric diag(1,1,…1,-1,-1,…,-1) where 1 and -1 occur p and q times, respectively).

The notion of subgroup is evident: Any subset of a group, which when equipped with the composition rule of the group is a group itself, is called a subgroup of the group. The tetragonal (dihedral) group $D_4$ is a subgroup of O, the group O is a subgroup of SO(3) and the group SO(3) a subgroup of SO(3,1). These examples leads to the chain of groups $D_4 \subset O \subset SO(3) \subset SO(3,1)$. Similarly, we have $SO(3) \subset SO(4) \subset SO(4,2)$.

### *3.2 Representations*

The concept of representation of a group is essential for applications. A linear representation of dimension m of a group G is an homomorphic image of G in GL(m,**C**), the group of regular complex mxm matrices. In other words, to each element of G there corresponds a matrix of GL(m,**C**) such that the composition rules for G and GL(m,**C**) are conserved by this correspondence. Among the various representations of a group, the unitary irreducible representations play a central rôle. From the mathematical viewpoint, a unitary irreducible representation (UIR) consist of unitary matrices which leave no space for the representation invariant. If invariant subspaces exist, the representation is said to be reducible. From the practical viewpoint, an UIR corresponds to a set of states (wavefunctions in the framework of quantum mechanics) which can be exchanged or transformed into each other *via* the elements of the groups. More intuitively, we may think of an UIR of dimension n as a set of n boxes into each of which it is possible to put an object or state, the n objects in the set presenting analogies or similar properties. This set of n boxes is often referred to as a n-plet. In Chemistry, we know that the UIR $T_1$ of dimension 3 of the group O is spanned by the orbitals $p_x$, $p_y$ and $p_z$. Mathematically, this means that any of these orbitals can be transformed into a linear combination of the 3 orbitals *via* the symmetry elements of O. In practice, the orbitals $p_x$, $p_y$ and $p_z$ are analogues ranged in the 3 boxes corresponding to the triplet $T_1$.

A finite group has a finite number of UIR's of finite dimension; the group O has five UIR's denoted $A_1$ (dimension 1), $A_2$ (dimension 1), E (dimension 2), $T_1$ (dimension 3) and $T_2$ (dimension 3) in Mulliken's notation. A compact group has a countable infinite number of UIR's of finite dimension; the group SO(3) has UIR's of dimensions 2l+1, denoted (l), where $l \in \mathbf{N}$, while the group SU(2) has UIR's of dimensions 2j+1, denoted (j), where $2j \in \mathbf{N}$. For a noncompact group, the UIR's are necessarily of infinite dimension; thus, the group SO(4,2) has an UIR that contains an infinite number of boxes.

As a consequence of the relation group–subgroup, each UIR of a given group is generally reducible as a sum of UIR's of the subgroup.

### *3.3 Examples*

We close this section with some examples of interest for what follows.

#### 3.3.1 The group SO(4)

The group SO(4) has UIR's denoted $(j_1,j_2)$ with $2j_1 \in \mathbf{N}$ and $2j_2 \in \mathbf{N}$ of dimension $(2j_1+1)(2j_2+1)$. The UIR (j,j) of SO(4) can be decomposed as (j,j)=(0)+(1)+…+(2j) in terms of UIR's of SO(3).

#### 3.3.2 The group SU(3)

The group SU(3) has UIR's noted (p,q) with $p \in \mathbf{N}$ and $q \in \mathbf{N}$ of dimension ½(p+1)(q+1)(p+q+2). For example, the following table lists a few UIR's relevant for the classification of particles.

| representation (p,q) | (0,0) | (1,0) | (0,1) | (1,1) | (3,0) | (0,3) |
|---|---|---|---|---|---|---|
| dimension of (p,q) | 1 | 3 | 3 | 8 | 10 | 10 |
| notation of the UIR of SU(3) | 1 | 3 | 3* | 8 or 8* | 10 | 10* |
| corresponding multiplet | singlet | triplet | anti-triplet | octet | decuplet | anti-decuplet |
| contents in UIR's of SU(2) | 1 = (1) | 3 = (½)+(0) | 3* = (½)+(0) | 8 = (0)+2 (½) +(1) | 10 = (0)+(½)+ (1)+($^3/_2$) | 10* = (0)+(½)+ (1)+($^3/_2$) |

#### 3.3.3 The group SO(4,2)

Finally, for the group SO(4,2), it is sufficient to say that there is an UIR, that we express here as h which may be decomposed as $h = \oplus_{\{j=0 \text{ to } \infty\}}$ (j,j) in terms of UIR's (j,j) of SO(4) and as $h = \oplus_{\{j=0 \text{ to } \infty\}} \oplus_{\{l=0 \text{ to } 2j\}}$ (l) in terms of UIR's (l) of SO(3). A careful examination of the last (Clebsch-Gordan) series shows that the UIR h of SO(4,2) can accommodate all the possible discrete quantum states of a hydrogen-like atom since we have

$$h = \oplus_{\{j=0 \text{ to } \infty\}} (j,j) \Rightarrow \Sigma_{\{j=0 \text{ to } \infty\}} (2j+1)^2 = \Sigma_{\{n=1 \text{ to } \infty\}} n^2$$

and

$$h = \oplus_{\{j=0 \text{ to } \infty\}} \oplus_{\{l=0 \text{ to } 2j\}} (l) \Rightarrow \Sigma_{\{n=1 \text{ to } \infty\}} \Sigma_{\{l=0 \text{ to } n-1\}} (2l+1)$$

in terms of dimensions.

## 4 Group theory and the periodic table of chemical elements

### *4.1 The importance of the atomic number Z*

The atomic number Z of an atom is of considerable importance for the periodic table. It determines the physico-chemical properties of the atom. It seems that such a result was published for the first time by van den Broek in 1913 although it was probably known at the same time by Rutherford. Of course, the number Z was not specified in the original Mendeleev chart but the ordering of the chemical elements in the chart reflected, with some exceptions, the increasing order of Z.[6] The work of Rutherford (1911) on the atomic nucleus and Moseley (1912) on X-ray spectroscopy showed that the atomic number Z of an element has a deep physical significance, namely the charge of its atomic nucleus expressed in units of minus the electronic charge. Consequently, Z turns out to be the number of electrons in the electronic cloud of the element in its neutral form. Moreover, according to the 1932 Heisenberg description of the nucleus in terms of protons and neutrons, the atomic number Z is also equal to the number of protons in the nucleus.

The first three quantitative manifestations of the atomic number Z concern: (i) the 1911 Rutherford formula for the differential cross-section in the scattering of alpha particles, (ii) the 1913 Moseley formula for the frequency of X-ray lines, and (iii) the 1913 Bohr-Balmer formula for the energy levels of a hydrogen-like atom. These formulae all contain Z and thus provide evidence that the atomic number completely characterizes a chemical element. However, it was only in 1923 that the "Commission Internationale des Poids et Mesures" recommended that a chemical element be defined by its atomic number Z or, equivalently, by its position in the Mendeleev periodic table.

### *4.2 The importance of quantum mechanics*

The periodicity, emphasized by Mendeleev in his table, of the chemical properties of elements corresponds in quantum mechanics to a repetition of the electronic structure of outer shells. In this connection, Bohr was the first to interpret the periodic system on the basis of the old quantum theory. Indeed, in 1922 he proposed a building-up principle for the atom based on a planetary (Bohr-Sommerfeld) model with elliptic orbits and on the filling of each orbit with a maximum of two electrons.[7] As a result, Bohr produced a periodic chart in which the transition metals and the rare-earth metals occupy a natural place. (Remember that the rare-earth metals occupy one single box in the conventional periodic table with 18 columns.) This new presentation of the periodic table led Bohr to suggest the existence of a new element with Z=72 and belonging to the family of Ti(Z=22) and Zr(Z=40). In fact, the element Hf(Z=72) was discovered in Zirconium minerals a short time after this suggestion and named Hafnium.

---

[6] Newlands used the term "atomic number" for the ordinal number specifying the position of an element in his classification.

[7] It is remarkable that Bohr's atomic "Aufbau" principle integrates: (i) The exclusion principle which was announced by Pauli in 1925 and (ii) The rule, credited to Madelung, which appeared in an explicit form in 1936.

Since Bohr's pioneer work, the periodic table has received other quantum-mechanical explanations. There are now several quantum-mechanical approaches to the periodic system of elements. In general, these approaches are based on an atomic-shell (Schrödinger or Dirac rather than Bohr) model, a given rule for filling up the various one-electron quantum levels, and the Pauli exclusion principle. As an illustration, the energy levels of a hydrogen-like atom define atomic shells nl, n'l', n"l", etc. Filling of these shells with Z electrons according to a given ordering rule and the Pauli exclusion principle produces a ground electronic configuration which characterizes the chemical element, of atomic number Z, in its neutral form. In the case of atomic shells resulting from self-consistent field calculations, the ordering rule directs the shells to fill up according to their increasing energies. In a somewhat idealized situation, the ordering rule is the so-called Madelung rule.

### *4.3 The Madelung rule*

The various shells nl and, more precisely, the various quantum states characterized by the triplets (n,l,m) can be obtained in the central-field model in (at least) two ways.

- One electron is embedded in the (spherically symmetric) Coulomb potential created by a nucleus of atomic charge Z: we thus have the familiar hydrogen atom (for Z=1) or an hydrogen-like atom (for Z arbitrary) for which n, l and m are the principal quantum number, orbital quantum number and magnetic quantum number, respectively. In this case, the one-electron energy E depends only on n and increases with n, a situation that we describe by E(n).
- Each electron of a complex atom is embedded in a spherically symmetric potential which corresponds to the superposition of the Coulomb field of the nucleus and an average (self-consistent) field created by the other electrons: l and m retain their usual significance while the number n is such that n-l-1 is the number of nodes of the radial wave function associated with the triplet (n,l,m). In this case, the one-electron energy E depends on the orbital quantum number l and on the number n. Two extreme situations may occur. First, E increases with n and, for a given value of n, with l, a situation we depict by E(n,l). Second, E increases with n+l and, for a given value of n+l, with n, a situation we depict by E(n+l,n).

In both cases, the spin quantum numbers s= ½ and $m_s$ (its projection on a z-axis) for each electron are introduced as further degrees of freedom. We are thus left with quantum states characterized by $(n,l,m,m_s)$.[8] A table of elements may be produced by filling the various quantum states with the Z electrons of an atom being distributed on the Z quantum states of lowest energy. The Pauli exclusion principle is taken into account by associating at most one electron to each quartet $(n,l,m,m_s)$ or two electrons to each triplet (n,l,m). The rule E(n) is of no use for the building-up of the neutral atoms. The rule E(n,l) can be used to construct the chemical elements in their neutral form from Z=1 to Z=18 and breaks down from Z=19. The latter fact was recognized by Bohr who replaced the rule E(n,l) by the rule E(n+l,n). The E(n+l,n) rule is referred to as the Madelung rule but may be also associated with Klechkovskii, Goudsmit or Bose.

[8] The possible values of n are n=1, 2, 3, …; for fixed n, the possible values of l are l=0, 1, …, n-1; for fixed l, the possible values of m are m=-l, -l+1, …, l; the possible values of $m_s$ are $m_s$ = - ½ and ½. Alternatively, we can use the quartet $(n,l,j,m_j)$ instead of $(n,l,m,m_s)$ where, for fixed l, we have j=|l-½|, l+½ and, for fixed j, $m_j$=-j, -j+1,…,j.

The Madelung rule, which was re-discovered on various occasions, has received much attention. In particular, it is well known that the statistical and semi-classical Thomas-Fermi model of the neutral atom can be used to account approximately for the first part of the Madelung rule. Furthermore, the first part of the rule E(n+l,n) has been investigated from an exactly soluble eigenvalue problem and from a group theoretical approach.

As a net result, the Madelung rule leads to the following order for the increasing one-electron energies:

| Electron states | Total degeneracy |
|---|---|
| (1s) | 2 |
| (2s, 2p) | 8 |
| (3s, 3p) | 8 |
| (4s, 3d, 4p) | 18 |
| (5s, 4d, 5p) | 18 |
| (6s, 4f, 5d, 6p) | 32 |
| (7s, 5f, 6d, 7p) | 32 |
| … | … |

where the levels in parentheses are also ordered after increasing energies. Note that the total degeneracy of the levels in parentheses reflects the variation $\Delta Z$ of the atomic number Z corresponding to two consecutive rare gases: $\Delta Z$(He $\rightarrow$ Ne) = $\Delta Z$(Ne $\rightarrow$ Ar) = 8, $\Delta Z$(Ar $\rightarrow$ Kr) = $\Delta Z$(Kr $\rightarrow$ Xe) = 18, $\Delta Z$(Xe $\rightarrow$ Rn) = $\Delta Z$(Rn $\rightarrow$ 118) = 32, etc.

We can then construct a periodic table from the Madelung rule in the following way. Let us consider the skeleton:

| | 0 | 1 | 2 | 3 | 4 | 5 | … |
|---|---|---|---|---|---|---|---|
| 1 | [1 1] | | | | | | |
| 2 | [2 2] | [3 2] | | | | | |
| 3 | [3 3] | [4 3] | [5 3] | | | | |
| 4 | [4 4] | [5 4] | [6 4] | [7 4] | | | |
| 5 | [5 5] | [6 5] | [7 5] | [8 5] | [9 5] | | |
| 6 | [6 6] | … | | | | | |
| … | | | | | | | |

where the rows are labelled with n = 1, 2, … and the columns with l = 0, 1, … and where the entry in the $n^{th}$ row and the $l^{th}$ column is [n+l n]. This provides us with the skeleton of a table where the various entries [n+l n] are filled in the "n+l n" dictionary order with chemical elements of increasing atomic numbers, the block {…} corresponding to [n+l n] containing 2(2l+1) elements. We thus obtain the following table where each element inside the block {…} is denoted by its atomic number Z:

| | | | | |
|---|---|---|---|---|
| {1 2} | | | | |
| {3 4} | {5 to 10} | | | |
| {11 12} | {13 to 18} | {21 to 30} | | |
| {19 20} | {31 to 36} | {39 to 48} | {57 to 70} | |
| {37 38} | {49 to 54} | {71 to 80} | {89 to 102} | … |
| {55 56} | {81 to 86} | {103 to 112} | {139 to 152} | … |
| {87 88} | {113 to 118} | {153 to 162} | … | |
| {119 120} | {163 to 168} | … | | |
| {169 170} | … | | | |
| … | | | | |

We are now ready to see how this table can be derived from group theory.

### *4.4 An SO(4,2)xSU(2) approach to the periodic table*

The periodic system of chemical elements has also been given a group theoretical articulation. Along these lines, a group structure of the periodic system was investigated by Barut on the basis of the group SO(4,2) and by Byakov, Kulakov, Rumer and Fet on the basis of the (direct product) group SO(4,2)xSU(2). In this connection, let us also mention the works by Odabasi, by Gurskii *et al.*, and by the present author.

As a preliminary point, we may ask the question "Why the group SO(4,2)xSU(2)?". First, it is clear that the groups SO(3) and SO(4) are subgroups of SO(4,2). The point symmetry group SO(3) describes the invariance under rotations in the three-dimensional space of any complex atom and, as shown by Pauli in 1926 and independently by Fock in 1935, the dynamical invariance group SO(4) explains the accidental degeneracy of the discrete spectrum of an hydrogen-like atom. Second, the continuous spectrum of an hydrogen-like atom is described by the dynamical invariance group SO(3,1), another subgroup of SO(4,2). Third, the dynamical non-invariance group SO(4,2) accounts for the discrete plus zero energy plus continuum states.[9] Fourth, the group SU(2) is a spectral group for labeling spin states. Consequently, SO(4,2) describes the dynamical part while SU(2) describes the spin part of the quantum states of a complex atom where the electronic correlation is not taken into account (the combination of SO(4,2) and SU(2) *via* a direct product indicates that the two parts are independent).

From a practical point of view, there is one representation of SO(4,2), viz., the representation described in the section on group theory, that contains all discrete states of the hydrogen atom: The set of all discrete states of the hydrogen atom can be regarded as spanning an infinite multiplet of SO(4,2). This is the starting point for the construction of a periodic chart based on SO(4,2)xSU(2). In such a construction, the chemical elements can be treated as structureless particles and/or considered as the various possible states of a dynamical system.

[9] It is perhaps worthy mentioning that, as shown by Cunningham in 1911, the group SO(4,2) leaves invariant the Maxwell equations. This makes sense if we remember that Maxwell equations constitute a (local) description of electromagnetism and that fundamentally Chemistry depends on electromagnetic interactions.

The relevant infinite multiplet of SO(4,2) can be organized into multiplets of SO(4) and SO(3). The multiplets of SO(4) may be disposed into rows characterized by the UIR labels n=1, 2, etc. The $n^{th}$ row contains n multiplets of SO(3) characterized by the UIR labels l=0, 1, …, n-1. This leads to a frame, with rows labeled by n and columns by l, in complete analogy with the skeleton of entries [n+l n] described for the Madelung rule. With the entry [n+l n] are associated 2l+1 boxes and each box can be divided into two boxes so that the entry [n+l n] contains 2(2l+1) boxes. The resultant doubling, i.e., 2l+1 $\rightarrow$ 2(2l+1), may be accounted for by the introduction of the group SU(2): We thus pass from SO(4,2) $\supset$ SO(4) $\supset$ SO(3) to SO(4,2)xSU(2) $\supset$ SO(4)xSU(2) $\supset$ SO(3)xSU(2). The 2(2l+1) boxes in the entry [n+l n] are organized into multiplets (j) of the group SU(2): For l different from 0, the entry contains two multiplets corresponding to j=l-½ and j= l+½ of lengths 2l and 2(l+½)+1=2(l+1), respectively; for l=0, the entry contains only one multiplet corresponding to j= ½ of length 2. Each box in the entry [n+l n] of the frame can be characterized by an address (n,l,j,$m_j$) with j and, for fixed j, $m_j$ increasing from left to right (for fixed j, the values of $m_j$ are $m_j$=-j, -j+1, …, j).

The connection with Chemistry is as follows. To each address (n,l,j,$m_j$) we can associate a value of Z. (The correspondence between address and atomic number shall be described in the second volume edited on the occasion of the 2003 Harry Wiener International Conference.) The box (n,l,j,$m_j$) is then filled with the chemical element of atomic number Z. This produces Table 1. It is remarkable that Table 1 is very close to the one arising from the Madelung rule. The only difference is that in Table 1 the elements in a given entry [n+l n] are arranged in one or two multiplets according to whether l is zero or different from zero.

Table 1 resembles, to some extent, other periodic tables. First, Table 1 is, up to the exchange of rows and columns, identical to the table introduced by Byakov, Kulakov, Rumer and Fet. The presentation adopted here, that follows a former work by the author, leads to a table the format of which is easily comparable to that of most tables in present day use. Second, Table 1 exhibits, up to a rearrangement, the same blocks as the table by Neubert based on the filling of only four Coulomb shells. Third, Table 1 presents, up to a rearrangement, some similarities with the table by Dash based on the principal quantum number, the law of second-order constant energy differences and the Coulomb-momentum interaction, except that in the Dash table the third transition group does not begin with Lu(Z=71), the lanthanide series does not run from La(Z=57) to Yb(Z=70) and the actinide series does not run from Ac(Z=89) to No(Z=102). Finally, it can be seen that Table 1 is very similar to the periodic table presented by Scerri at the 2003 Harry Wiener International Conference. As a point of fact, to pass from the format of Table 1 to the format of the Scerri table, it is sufficient to do a symmetry with respect to an axis parallel to the columns and located at the left of Table 1 and then to get down the p-blocks (for l=1) by one unit, the d-blocks (for l=2) by two units, etc.

| | | | | | | | | | | | | | | | | | | | | | | | | | | | | | | | | |
|---|---|---|---|---|---|---|---|---|---|---|---|---|---|---|---|---|---|---|---|---|---|---|---|---|---|---|---|---|---|---|---|---|
| 1 | 2 | | | | | | | | | | | | | | | | | | | | | | | | | | | | | | | |
| H | He | | | | | | | | | | | | | | | | | | | | | | | | | | | | | | | |
| 3 | 4 | 5 | 6 | 7 | 8 | 9 | 10 | | | | | | | | | | | | | | | | | | | | | | | | | |
| Li | Be | B | C | N | O | F | Ne | | | | | | | | | | | | | | | | | | | | | | | | | |
| 11 | 12 | 13 | 14 | 15 | 16 | 17 | 18 | 21 | 22 | 23 | 24 | 25 | 26 | 27 | 28 | 29 | 30 | | | | | | | | | | | | | | | |
| Na | Mg | Al | Si | P | S | Cl | A | Sc | Ti | V | Cr | Mn | Fe | Co | Ni | Cu | Zn | | | | | | | | | | | | | | | |
| 19 | 20 | 31 | 32 | 33 | 34 | 35 | 36 | 39 | 40 | 41 | 42 | 43 | 44 | 45 | 46 | 47 | 48 | 57 | 58 | 59 | 60 | 61 | 62 | 63 | 64 | 65 | 66 | 67 | 68 | 69 | 70 | |
| K | Ca | Ga | Ge | As | Se | Br | Kr | Y | Zr | Nb | Mo | Tc | Ru | Rh | Pd | Ag | Cd | La | Ce | Pr | Nd | Pm | Sm | Eu | Gd | Tb | Dy | Ho | Er | Tm | Yb | |
| 37 | 38 | 49 | 50 | 51 | 52 | 53 | 54 | 71 | 72 | 73 | 74 | 75 | 76 | 77 | 78 | 79 | 80 | 89 | 90 | 91 | 92 | 93 | 94 | 95 | 96 | 97 | 98 | 99 | 100 | 101 | 102 | 121-138 |
| Rb | Sr | In | Sn | Sb | Te | I | Xe | Lu | Hf | Ta | W | Re | Os | Ir | Pt | Au | Hg | Ac | Th | Pa | U | Np | Pu | Am | Cm | Bk | Cf | Es | Fm | Md | No | no |
| 55 | 56 | 81 | 82 | 83 | 84 | 85 | 86 | 103 | 104 | 105 | 106 | 107 | 108 | 109 | 110 | 111 | 112 | 139 | 140 | 141 | 142 | 143 | 144 | 145 | 146 | 147 | 148 | 149 | 150 | 151 | 152 | … |
| Cs | Ba | Tl | Pb | Bi | Po | At | Rn | Lr | Rf | Db | Sg | Bh | Hs | Mt | X? | X? | X? | no | no | no | no | no | no | no | no | no | no | no | no | no | no | … |
| 87 | 88 | 113 | 114 | 115 | 116 | 117 | 118 | 153 | 154 | 155 | 156 | 157 | 158 | 159 | 160 | 161 | 162 | … | | | | | | | | | | | | | | |
| Fr | Ra | no | X? | no | X? | no | no | no | no | no | no | no | no | no | no | no | no | … | | | | | | | | | | | | | | |
| ... | | | | | | | | | | | | | | | | | | | | | | | | | | | | | | | | |
| ... | | | | | | | | | | | | | | | | | | | | | | | | | | | | | | | | |

**Table 1: The *à la* SO(4,2)xSU(2) periodic table (X? = observed but not named, no = not observed)**

The main distinguishing features of Table 1 may be seen to be the following: (i) Hydrogen is in the family of alkali metals, (ii) Helium belongs to the family of alkaline earth metals, and (iii) the inner transition series (lanthanides and actinides) as well as the transition series (iron group, palladium group and platinum group) occupy a natural place in the table. This contrasts with the conventional tables in 8(9) or 18 columns where: (i) Hydrogen is sometimes located in the family of halogens, (ii) Helium generally belongs to the family of noble gases, and (iii) the lanthanide series and the actinide series are generally treated as appendages. Furthermore, in Table 1 the distribution in of the elements Z=104 to 120 is in agreement with Seaborg's predictions based on atomic-shell calculations. In contrast to his predictions, however, Table 1 shows that the elements Z=121 to 138 form a new family having no homologue among the known elements. In addition, Table 1 suggests that the family of super-actinides contains the elements Z=139 to 152 (and not Z=122 to 153 as predicted by Seaborg). Finally, note that the number of elements afforded by Table 1 is *a priori* infinite.

Another important peculiarity of Table 1 is the division, for l different from zero, of the l-block into two sub-blocks of length 2l and 2(l+1). As an illustration, we get two sub-blocks of length 4 and 6 for the d-blocks (corresponding to l=2) and two sub-blocks of length 6 and 8 for the f-blocks (corresponding to l=3). In the case of rare earth elements, the division into light or ceric rare earth elements and heavy or yttric rare earth elements is quite well known. It has been underlined on the basis of differences for the solubilities of some mineral salts and on the basis physical properties as, for example, magnetic and spectroscopic properties of rare earth compounds. The division for the rare earth elements and the other l-blocks (with l different from 0) was explained by using the Dirac theory for the electron. In particular, it was shown that the calculation of magnetic moments *via* this theory accounts for the division of the rare earth elements. It is to be noted that the valence 2 of the Samarium ion can be explained by the division of the f-block into two sub-blocks: in a shell-model picture, the ion $Sm^{2+}$ then corresponds to a completely filled shell (with six electrons) associated with the first sub-block.

### *4.5 Towards a quantitative approach*

The application of the group SO(4,2)xSU(2) has up to now been limited to qualitative aspects of the periodic table. We would like to present here in outline a program (inherited from nuclear physics and particle physics) for using SO(4,2)xSU(2) in a quantitative way. A detailed treatment will be presented in a subsequent paper.

The group SO(4,2), locally isomorphic to the special unitary group in 2+2 dimensions SU(2,2), is a Lie group of order=15 and rank=3. It has therefore 15 generators involving 3 Cartan generators (i.e., generators commuting between themselves). In addition, it has 3 invariants or Casimir operators (i.e., independent polynomials in the generators that commute with all generators of the group). Therefore, we have a set of 3+3=6 operators that commute between themselves. Indeed, this set is not complete from the mathematical point of view. According to a lemma by Racah, we need to find ½(order–3rank) = 3 additional operators in order to get a complete set. It is thus possible to build a complete set of 6+3=9 commuting operators. Each of the 9 operators can be taken to be self-adjoint and thus, from the quantum-mechanical point of view, can describe an observable.

The next step is to connect chemical and physical properties (like ionization potentials, melting temperatures, atomic volumes, densities, magnetic susceptibilities, solubilities, etc.) to the 9 invariant operators. In most cases, this can be done by expressing a chemical observable associated with a given property as a linear combination of the 9 invariant operators.

The last step is to fit the various linear combinations to experimental data. For each property this will lead to a formula or phenomenological law that can be used in turn for making predictions concerning the chemical elements for which no data are available. In this respect, the approach described above resembles the topological one presented by G. Restrepo at the Harry Wiener Conference.

## 5 A periodic table in the sub-atomic word

### *5.1 The use of group theory for particles and their interactions*

The use of group theory in particle physics differs from the one in Chemistry in the sense that for particles and their interactions we have to distinguish between external and internal symmetries. The external symmetries take place in the Minkowski space-time and are described by the Lorentz group and the Poincaré group. The internal symmetries are described by classification or flavor groups and describe intrinsic properties of particles (like isospin, strangeness, etc.). The two kinds of symmetries occur in the concept of gauge group: The interaction between particles is deduced from a (local invariance) gauge principle which gives rise to the mediating or gauge fields that transmit the interaction.

The classification groups are postulated by looking at conservation laws, symmetries and regularities. For n flavors of quarks ($n \leq 6$), the best classification group is $SU(n)_f$ (or $SU(2n)_f \supset SU(2)xSU(n)_f$ if mutiplets involving particles of different spins are involved). The typical gauge groups are the color group $SU(3)_c$ for the strong interactions and SU(2)xU(1) for electromagnetic and weak interactions. These groups are nonabelian (or non commutative) groups so that they yield more than one gauge particle in contradistinction with the abelian gauge group U(1) of electromagnetism which gives rise to one gauge particle (the photon). Let us also mention the interest of the gauge groups SU(5), SO(10), $E_6$, $E_7$ and $E_8$ for the unification of strong and electro-weak interactions.

In this paper, we shall limit ourselves to the flavor group $SU(3)_f$ for 3 flavors of quarks and to the groups $SU(3)_c$ and SU(2)xU(1) for the description of the strong and electro-weak interactions, respectively.

### *5.2 The standard model*

The standard model for particles and their interactions goes back to the 60’s and was mainly developed by Glashow, Weinberg and Salam. The last experimental evidence for this model was the discovery of the tau-neutrino in 2000. In the current formulation, the SU(3)xSU(2)xU(1) standard model involves 12 matter particles (fermions) and 12 mediating particles (bosons) for describing the strong, electromagnetic and weak interactions. Indeed, the electromagnetic and weak interactions are unified *via* an SU(2)xU(1) gauge group (spontaneously broken according to SU(2)xU(1) $\supset$ U(1)) and the strong interactions are described by an SU(3) gauge or color group. To these 12+12=24 particles (plus the anti-particles associated with the 12 fermions), we must add the Higgs (or feeding) particle, a particle of charge 0 (neutral particle) and spin 0 (scalar boson) required to account for the mass of massive particles. We give below tables for the matter, mediating and feeding particles.

| | | | | |
|---|---|---|---|---|
| matter particles of generation 1 | (up) quark u | (down) quark d | electron | e-neutrino |
| matter particles of generation 2 | (charm) quark c | (strange) quark s | muon | mu-neutrino |
| matter particles of generation 3 | (top or truth) quark t | (beauty) quark b | tau | tau-neutrino |
| spin | 1/2 | 1/2 | 1/2 | 1/2 |
| electric charge | 2/3 | -1/3 | -1 | 0 |

| mediating particle | photon | boson $W^+$ | boson $W^-$ | boson $Z^0$ | 8 gluons |
|---|---|---|---|---|---|
| mass | massless | massive | massive | massive | massless |
| spin | 1 | 1 | 1 | 1 | 1 |
| electric charge | 0 | 1 | -1 | 0 | 0 |
| color charge | no | no | no | no | yes |
| interaction | electro-weak: unification of two interactions | | | | strong |
| gauge group | SU(2)xU(1) $\supset$ U(1) | | | | SU(3) |

| feeding particle | Higgs boson |
|---|---|
| spin | 0 |
| electric charge | 0 |

We note that the table for the matter particles is in some sense a periodic table with 3 periods (or generations) ranged in lines and 4 families ranged in columns (the up quark family, the down quark family, the electron family and the neutrino family). Each family corresponds to a given electric charge.

We shall see in the remaining part of this paper how we arrive at the last three tables starting from 1897 with the discovery of the electron by Thompson.

### *5.3 How has the periodic table of particles arisen?*

#### 5.3.1 1932: The ideal world (the Golden Age)

The situation was very simple in 1932: Two leptons (electron and neutrino) and two hadrons (proton and neutron) were sufficient for understanding matter. The electron was made evident by Thomson in 1897 and the neutrino (or electron-neutrino) was predicted[10] by Pauli in 1930 in order to explain the continuous spectrum of the energy of electrons in natural β-radioactivity. The proton,[11] a constituent of the nucleus, was proved to exist by Rutherford in 1909 and the neutron, the other constituent of the nucleus, was discovered in cosmic rays by Chadwick in 1932. To these four particles of spin ½, we must add the corresponding anti-particles[12] (of spin ½) and the light quantum or photon[13]of spin 1. The anti-particle of the electron, the positron, was discovered in cosmic rays by Anderson in 1933 but the other anti-particles were not observed at this time.

The symmetry between the leptonic world (electron and neutrino) and the hadronic world (proton and neutron) is total. In 1932, the two leptons and two hadrons occur as the sole universal constituents of matter. The photon has then an intermediate status between matter and energy.

The similarity, as far as the electromagnetic properties are not concerned, of the properties between proton and neutron (same spin, very close masses, similar strong interactions) led Heisenberg in 1932 to develop a theory of the nucleus based on the group SU(2). In this theory, proton and neutron are two possible states of a same entity, the nucleon. They are accommodated in the UIR 2 of SU(2), corresponding to a new quantum number, the isospin equal to ½. The two states are distinguished by their z-component of the isospin quantum number (+ ½ for the proton and – ½ for the neutron). The group SU(2) is thus the first classification of flavor group used in sub-atomic physics.

The Heisenberg theory was followed in 1933 by the Fermi theory of β-decay that remained for a long time, with the V-A theory, the foundation for the current theory of weak interactions and in 1935 by the Yukawa theory (with the prediction of the Yukawa particle) that heralded the current theory of strong interactions.

### 5.3.2 1962: The world gets complicated

From 1932 to 1962, the number of particles continued to grow. One more lepton – an heavy electron called muon – was discovered in 1937 in cosmic rays and many hadrons (baryons and mesons) were discovered in cosmic rays or produced in accelerators. In particular from 1947 to 1953, pions,[14] kaons and hyperons were observed in cosmic rays.

---

[10] This is an example of eka-process in nuclear physics. The neutrino, called neutron by Pauli and re-baptized as neutrino by Fermi in 1933, was postulated together with some of its properties in order not to violate the principle of conservation of energy-momentum and to obey spin statistics. It (actually its anti-particle) was observed by Reines and Cowan in 1956.

[11] Rutherford proposed in 1920 to call proton the nucleus of the hydrogen atom.

[12] The notion of anti-particle was introduced by Dirac in 1928 as a sub-product of its celebrated relativistic equation. A particle and its anti-particle have the same mass and same vector quantum numbers but opposite charges and opposite additive quantum numbers.

[13] The name photon was given to the light particle at the Solvay Congress in Brussels in 1927.

[14] As another example of the eka-process, the particle (meson) predicted by Yukawa in 1935 was observed as the charged pions $\pi^+$ and $\pi^-$ in 1947 and the neutral pion $\pi^0$ in 1949.

At the beginning of the 50's, after the discovery of strange[15] particles (kaons and hyperons), the situation in particle physics was very similar to that preceding the introduction of the periodic table of chemical elements. The time was ripe for developing a classification of particles on the basis of analogies, conservation laws and symmetries. The new particles and resonances were first classified into multiplets of the group SU(2). For example, the singlet ($\Lambda^0$), the doublet ($K^0$,$K^+$) and the triplet ($\Sigma^-$,$\Sigma^0$,$\Sigma^+$) were accommodated into the UIR of dimension 1, 2 and 3 of SU(2), respectively. However, it soon appeared that the group SU(2) was not sufficient for an efficient classification. Following the pioneer work by Sakata in 1953, Gell-Mann and Ne'eman developed independently in 1961 a revisited version of the flavor group SU(3) for the classification of hadrons. In this classification, called the eightfold way,[16] the hadrons with similar properties (same spin, same parity, similar masses, etc.) , are arranged in the same UIR or into a direct sum of UIR's of SU(3). As an illustration, the baryons $(½)^+$, with spin ½ and parity +1, are put in the UIR 8 of SU(3) and the pseudo-scalar mesons ($0^-$), with spin 0 and parity –1, are put in the representation 8+1 of SU(3).

The group SU(3) was thus useful not only for classification purposes (qualitative aspects) but also for both quantitative (mainly calculation of cross sections and derivation of mass formulas) and predictive purposes. In this regard, the celebrated prediction by Gell-Mann (probably foreseen as early as 1956) of the hyperon $\Omega^-$ is another example of the eka-process in high energy physics. Indeed in 1962, nine baryons of spin 3/2 and parity +1 were known and could be accommodated in the UIR 10 of SU(3) thus leaving an empty box as shown in the following figure.

| | | | |
|---|---|---|---|
| $\Delta^-$ | $\Delta^0$ | $\Delta^+$ | $\Delta^{++}$ |
| $\Sigma^{*-}$ | $\Sigma^{*0}$ | $\Sigma^{*+}$ | |
| $\Xi^{*-}$ | $\Xi^{*+}$ | | |
| empty box | | | |

Gell-Mann was able to predict in 1962 that the empty box should be occupied with a particle of spin 3/2, parity –1, charge –1. He also gave some decay modes for this particle and, owing to the Gell-Mann/Nishijima mass formula, predicted a mass of 1672 MeV/$c^2$. The corresponding particle $\Omega^-$ was observed in bubble chambers at Brookhaven in 1964 with the predicted specifications.

To sum up the situation around 1962, the following table gives the classification of the baryons $(½)^+$ and $(3/2)^+$, the pseudo-scalar mesons $(0)^-$ and the vector mesons $(1)^-$. (In the table, $\underline{A}$ stands for the anti-particle of the particle A.)

[15] The word "strange" was introduced for emphasizing that the strange particles decay through weak interactions although they undergo strong interactions.

[16] The name was given with reference to a Chinese aphorism (according to Buddha, 8 good actions render the life easier). In addition, the group SU(3) has 8 generators and it took 8 years after the first trial by Sakata to develop the eightfold way.

| | quartet of SU(2) | triplets of SU(2) | doublets of SU(2) | singlets of SU(2) |
|---|---|---|---|---|
| octet $(1/2)^+$: 8 baryons in the UIR 8 of SU(3) | | $\Sigma^-, \Sigma^0, \Sigma^+$ | n, p<br>$\Xi^-, \Xi^0$ | $\Lambda^0$ |
| decuplet $(3/2)^+$: 10 baryons in the UIR 10 of SU(3) | $\Delta^-, \Delta^0, \Delta^+, \Delta^{++}$ | $\Sigma^{*-}, \Sigma^{*0}, \Sigma^{*+}$ | $\Xi^{*-}, \Xi^{*0}$ | $\Omega^-$ |
| nonet $(0)^-$: 9 mesons in the sum 8+1 of SU(3) | | $\pi^-, \pi^0, \pi^+$ | $K^0, K^+$<br>$\underline{K}^+, \underline{K}^0$ | $\eta^0$<br>$\eta'^0$ |
| nonet $(1)^-$: 9 mesons in the sum 8+1 of SU(3) | | $\rho^-, \rho^0, \rho^+$, | $K^{*0}, K^{*+}$<br>$\underline{K}^{*+}, \underline{K}^{*0}$ | $\omega^0$<br>$\phi^0$ |

### 5.3.3 1964: The world gets simpler

The multiplication of particles (~100 particles and resonances in 1964) was a good reason for trying to simplify the classification scheme afforded by SU(3). As a notable extension, the spin dependence was included *via* the introduction of an SU(2) group leading to the SU(6) ⊃ SU(2)xSU(3) model. In this model, all the pseudo-scalar mesons $(0)^-$ and vector mesons $(1)^-$ (a total of 9 + 3x9 = 36 states including spin) were classified in the UIR 36 of SU(6); similarly the baryons $(½)^+$ and $(3/2)^+$ (a total of 2x8 + 4x10 = 36 states including spin) were classified in the UIR 56 of SU(6). From a quantitative point of view, the flavor group SU(6) proved to be useful for calculating the ratio of the magnetic moments of the nucleons p and n.

The biggest success of SU(3), and consequently SU(6), came from the introduction of quarks. Gell-Mann and Zweig noticed independently in 1964 that the fundamental UIR's 3 and 3* of SU(3) were not occupied with particles. Therefore, they introduced 3 particles with fractional charges (named quarks by Gell-Mann and aces by Zweig) in the UIR 3 and their 3 anti-particles (anti-quarks or anti-aces) in the UIR 3*. The main specifications of these particles and anti-particles are given in the following table.

| | baryonic number | electric charge | hypercharge number | isospin number | z-component of isospin | strangeness number |
|---|---|---|---|---|---|---|
| u | 1/3 | 2/3 | 1/3 | 1/2 | 1/2 | 0 |
| d | 1/3 | -1/3 | 1/3 | 1/2 | -1/2 | 0 |
| s | 1/3 | -1/3 | -2/3 | 0 | 0 | -1 |
| $\underline{u}$ | -1/3 | -2/3 | -1/3 | 1/2 | -1/2 | 0 |
| $\underline{d}$ | -1/3 | 1/3 | -1/3 | 1/2 | 1/2 | 0 |
| $\underline{s}$ | -1/3 | 1/3 | 2/3 | 0 | 0 | 1 |

The next idea was to consider these particles and anti-particles as fundamental particles for constructing all hadrons. In other words, every baryon or meson is made of quarks and anti-quarks. More specifically[17]

| |
|---|
| baryon = bound state of three quarks (q q q) |
| anti-baryon = bound state of three anti-quarks ($\underline{q}$ $\underline{q}$ $\underline{q}$) |
| meson or anti-meson = bound state of one quark and one anti-quark (q $\underline{q}$) |

The justification is founded on the fact that the charge and all additive (in a scalar or vector form) quantum numbers of a given hadron can be deduced from the corresponding quantities of its constituent quarks and/or anti-quarks. As an example, the proton

p = u u d

is made of two quarks up and one quark down so that its baryonic number is 1/3+1/3+1/3=1, its charge is 2/3+2/3-1/3=1, its spin and isospin quantum numbers are the result of the coupling of three angular momenta ½. Similarly, the pion

$\pi^+ = u\,\underline{d}$

is made of one quark u and one anti-quark $\underline{d}$ and it can be verified that all its quantum numbers may be deduced from the ones of u and $\underline{d}$.

The quark model constitutes more than a simple book-keeping for conveniently summarizing the intrinsic quantum numbers of the strongly interacting particles. Although it was used for (re-)deriving in a simple way many results about static properties of hadrons (e.g., the ratio of the magnetic moments for p and n), a major problem occurred very soon after its introduction. To understand the problem, consider the particle $\Omega^-$. In the quark model, we should have

$\Omega^- = s\,s\,s$

and, as an immediate consequence, the interchange of two strange quarks in the wavefunction for $\Omega^-$ does not change its sign as it should since $\Omega^-$ is a fermion. At first sight, the Pauli exclusion principle is violated. Several remedies (new statistics or introduction of color) were considered to overcome this difficulty. With the introduction of a new quantum number called color,[18] each quark may exist in three different states (or colors): R(red), G(green) and B(blue) and in the composition rule for hadrons only the anti-symmetric (for baryons) and symmetric (for mesons) bound states have to be considered, e.g.,

[17] In an abbreviated form, we can write baryon = $q^3$, anti-baryon = $\underline{q}^3$ and meson = $q\underline{q}$. Recently, indication of the existence of a new meson (actually a molecule $q^2\underline{q}^2$ of two mesons) and a new baryon (actually a pentaquark state $q^4\underline{q}$) have been reported.

[18] The imaged notion of flavor and color of quarks (and hadrons) should not be taken too seriously. The flavor of a quark indicates its type and, for a given type, its color refers to a further quantum number. At present, six flavors are known (u,d,c,s,t,b) and each flavor may come in three colors (R,G,B).

$\Omega^- = s_R\, s_G\, s_B + s_G\, s_B\, s_R + s_B\, s_R\, s_G - s_R\, s_B\, s_G - s_B\, s_G\, s_R - s_G\, s_R\, s_B$
$p = u_R\, u_G\, d_B + u_G\, u_B\, d_R + u_B\, u_R\, d_G - u_R\, u_B\, d_G - u_B\, u_G\, d_R - u_G\, u_R\, d_B$
$\pi^+ = u_R\, \underline{d}_{\underline{R}} + u_G\, \underline{d}_{\underline{G}} + u_B\, \underline{d}_{\underline{B}}$

To go further, each quark q is associated with a triplet of SU(3) containing $q_R$, $q_G$ and $q_B$ and with the UIR 3 of a new SU(3) group called the color group $SU(3)_c$. A similar association holds for $\underline{q}_{\underline{R}}$, $\underline{q}_{\underline{G}}$ and $\underline{q}_{\underline{B}}$ (with anti-colors $\underline{R}$, $\underline{G}$ and $\underline{B}$) and the UIR 3* of $SU(3)_c$.

The consequence of introducing of $SU(3)_c$ is at the root of a theory (QCD) for describing strong interactions. Indeed, $SU(3)_c$ is a gauge group and the bosons for mediating interactions between quarks are the 8 gluons, corresponding to the 8 independent colored combinations

(3 colors) x (3 anti-colors) – 1 = 9 – 1 = 8

which span the regular UIR 8 of $SU(3)_c$ (the term –1 corresponds to the white combination $R\underline{R}+G\underline{G}+B\underline{B}$). Thus the quarks interact by exchanging gluons and the strong interactions between hadrons can be considered as a residual form of the color force.[19] The gluons are massless particles, a result that reflects the fact that $SU(3)_c$ is an unbroken symmetry. By contrast, the flavor group $SU(3)_f$ (i.e., the group SU(3) introduced in Section 5.3.2) corresponds to a broken symmetry and the particles in a given UIR of $SU(3)_f$ may have different nonzero masses.

### 5.3.4 Other important steps

At this stage, it is perhaps interesting to give without detail some of the important steps which led to the tables of the standard model given in Section 5.2. The indicated dates are often benchmarks corresponding to the center of gravity of periods during which decisive results were obtained. In fact, the standard model of elementary particles and their interactions, based on $SU(3)_c$xSU(2)xU(1), was mainly developed between 1961 and 1973.

- 1965: introduction of Quantum Chromo-Dynamics (QCD), a gauge theory based on the group $SU(3)_c$ of the strong interaction.[20] The $SU(3)_c$ model corresponds to an exact symmetry yielding 8 massless gauge bosons, the gluons. In this model, each hadron is made of quarks (valence and see quarks) and a soup of gluons.
- 1968: unification of weak and electromagnetic interactions, a gauge theory based on the chain of groups SU(2)xU(1) $\supset$ U(1), the queue group U(1) indicating that the SU(2)xU(1) symmetry is spontaneously broken. The two main predictions of the SU(2)xU(1) ⊃ U(1) electro-weak model, mainly developed by Glashow, Salam, Ward and Weinberg, were the so-called neutral currents and the intermediate bosons ($W^-$, $W^+$, $Z^0$) observed at CERN in 1973 and 1983, respectively.

[19] The situation is very similar in molecular physics where the van der Waals forces between molecules may be considered as a residual form of the electromagnetic forces between charged particles of the constituent atoms.

[20] QCD is to strong interactions what QED (Quantum Electro-Dynamics), another gauge theory, is to electromagnetic interactions. Both QCD and QED are relativistic quantum field theories. The interaction between particles takes place by the exchange of photons (which do not carry an electric charge) in QED and of gluons (which do carry color charges) in QCD.

- 1970: prediction of a fourth quark (the charm quark c) to explain that certain decays of particles, *a priori* permitted, are not observed. The charm quark was made evident with the discovery of the $J/\Psi = c\underline{c}$ resonance in 1974. Note that, with the introduction of a fourth quark (referred to as the November revolution), the flavor group $SU(3)_f$ has to be replaced by an $SU(4)_f$ flavor group but the color group remains the $SU(3)_c$ dynamics group.
- 1962-2000: discovery of various leptons and quarks as shown below[21]

| particle | mu-neutrino | tau | tau-neutrino | quark b | quark t |
|---|---|---|---|---|---|
| discovered in | 1962 | 1975 | 2000 | 1977 | 1995 |
| predicted in | | | | 1974 | 1977 |

thus completing the periodic table for matter particles of Section 5.2.

### *5.4 Some current pivotal research in particle physics*

To close this paper, we give briefly a nonexhaustive list of some directions being taken in particle physics research.

#### 5.4.1 Theoretical aspects

- Grand unified theories. The unification of strong and electro-weak interactions and the search of composite particles (called preons, rishons or quips in some former works) of quarks and leptons has been addressed by means of group theory. The first grand unified model (developed by Georgi and Glashow in 1974) was based on the group SU(5). Other models have been introduced on the basis of the group SO(10) and of the exceptional groups $E_6$, $E_7$ and $E_8$. As a general trend, all these models lead to the prediction of monopoles and the decay of the proton. There is up to now no experimental evidence of monopoles and the lower limit for the proton lifetime is $10^{32}$ years.[22]

- Supersymmetry. This research direction concerns the unification of the matter, mediating and Higgs fields. Supersymmetry is a (postulated) symmetry between bosons and fermions. It takes its origin in an extension of the Poincaré group introduced in order to unify internal and external symmetries in a nontrivial manner. As a consequence of supersymmetry, to each existing particle is associated a new particle called supersymmetric partner or superparticle. If the particle is a fermion, the superparticle is said to be a sfermion and if the particle is a boson, the superparticle is said to be a bosino. As an illustration, we have the three following tables.

---

[21] With 6 quarks and 6 leptons, the symmetry between quarks and leptons is total.

[22] A possible decay mode is $p \to e^+ + \pi^0$. With the decay of a proton in an hydrogen atom (one proton and one electron) everything ends up as light since $\pi^0 \to 2\gamma$ and $e^- + e^+ \to 2\gamma$.

| | | |
|---|---|---|
| fermion (spin ½) | ↔ | sfermion (spin 0) |
| quark | ↔ | squark |
| electron | ↔ | selectron |
| muon | ↔ | smuon |
| tau | ↔ | stau |
| boson (spin 1) | ↔ | bosino (spin ½) |
| photon | ↔ | photino |
| gluon | ↔ | gluino |
| W | ↔ | wino |
| Z | ↔ | zino |
| boson (spin 0) | ↔ | bosino (spin ½) |
| Higgs | ↔ | higgsino |

- Structure of space-time. The introduction of additional dimensions to the usual 3+1 space-time dimensions yields the introduction of new objects (strings and branes) which are alternative to the point particles. The M-theory (in 11-dimensional spaces) introduced in 1995 constitutes, through the idea of duality, an extension of superstring theory.

### 5.4.2 Experimental aspects

Only the briefest of details are given here.

- Evidence for extra dimensions.
- Signatures of CP (charge-parity symmetry) violation.
- The quest for dark matter.
- Neutrino oscillations.
- Quark-gluon plasma.
- The quest for the Higgs particle.
- The quest for the W’ and Z’ bosons.
- The quest for superparticles.

All the above-listed fields are presently the object of important experimental developments at DESY and GSI in Germany, Fermilab and SLAC in the U.S.A., Kamiokande in Japan, SNO in Canada and CERN & Gran Sasso in France-Switzerland-Italy. The LHC (Large Hadron Collider) at CERN should be operative in 2007 and is aimed at the last four items.

## 6 Closing remarks

The historical outlines in this paper highlight two distinctive facets of science, *viz.*, that progress is achieved through a never-ending cycle of experiment-prediction-experiment-etc. and that the resulting discoveries and developments generally arise from the work of many people. These facets are well illustrated by the eka-process in both Chemistry and Physics with the numerous Mendeleev analogues and the work by Pauli, Yukawa, Gell-Mann and others concerning nuclei and fundamental particles.

The present work in this paper focuses on the relevance and interest of group theory for atoms and for particles and fields. In these two complementary areas, group theory is used as a tool, together with other tools,[23] for developing a model such as the SO(4,2)xSU(2) model for the chemical elements and the SU(3)xSU(2)xU(1) model for particles.

For the chemical elements, the SO(4,2)xSU(2) approach based on a joint use of group theory and quantum mechanics produces a periodic table in agreement with the phenomenological Madelung rule. In addition, this approach allows for the existence of two sub-blocks in a given l-block. Thus, the group SO(4,2)xSU(2) turns out to be a very useful group for classification purposes. As suggested in this work, it has also serious potentialities for describing chemical and physical properties.

In the SU(3)xSU(2)xU(1) model of the sub-atomic world, the eka-process often appears to be of a purely mathematical nature. For example, the quarks and gluons were initially just mathematical entities satisfying conservation laws related to internal symmetries before they acquired the status of physical object.[24] Much the same may be said of the Higgs mechanism before being considered as giving rise to a (still unobserved) particle.

The periodic table of quarks and leptons, raises some interesting questions. Why 3 generations? Only the u and d quarks are stable and all every-day life matter on earth is made of particles of the first generation. What is the rôle of the other quarks?

In conclusion, it is interesting to note that the SO(4,2)xSU(2) periodic table for chemical elements is infinite (thereby offering some hope of finding new stability islands for super-heavy nuclei) and that the periodic table for quarks and leptons is presently finite (with 3 generations, without excluding the possibility to have other generations at higher energy).

---

[23] As, for instance, nonrelativistic or relativistic quantum mechanics for chemical elements, relativistic quantum field theory and gauge theory for particles.

[24] The quarks and gluons are confined inside the hadrons (with an "asymptotic or ultra-violet freedom" at high energy and an "infra-red slavery" at low energy). They cannot be isolated from the hadrons: The hadrons are white (white = superposition of 3 colors) in imaged terms of color. However, indirect evidence for quarks and gluons were found at the end of the 60's for quarks (with experiments on deep inelastic scattering of electrons on protons, showing the existence of the Feynman partons inside the proton) and in 1979 for gluons (with the discovery of 3-jet structures).

**Acknowledgments**
I would like to thank the Wiener family and the organizers of the Second Harry Wiener International Memorial Conference, especially Dennis Rouvray (Conference President), Bruce King (Conference Secretary) and Michael Wingham (Conference Vice-President) to have made possible this beautiful interdisciplinary meeting.

I am indebted to the late Georges Kayas who kindly sent me his nice essay on the history of particles, to Charles Ruhla who kindly put at my disposal some of his lecture notes and to Ray Hefferlin who stimulated my interest in SO(4,2)xSU(2) for the classification of atoms and molecules. Thanks are also due to Françoise Gaume, Xavier Oudet, Georges Lochak and Eric Scerri for useful correspondence and to Tidjani Négadi, Jean-Claude Poizat, Jean-Paul Martin, Maurice Giffon, Aldo Deandrea, Dany Davesne and Mariasusai Antony for interesting discussions. Finally, I should like to thank Michael Cox for his helpful editorial assistance and suggestions with the final polish.

This paper is dedicated to the memory of Jean Gréa who, like Harry Wiener, was a universal man. Jean was a physicist with a large flow of activities in theoretical physics. He started with nuclear physics, then switched to pre-quantum mechanics (theory of hidden variables), continued with simulations in physics and the theory of signal and finally made important contributions in didactics and philosophy of sciences. He was the co-founder of the LIRDHIST (*Laboratoire Interdisciplinaire de Recherche en Didactique et en Histoire des Sciences et Techniques*) at the *Université Claude Bernard Lyon 1*. As a teacher, he knew how to communicate his enthusiasm to students. He was very much oriented to other people. He knew how to listen to others, how to answer questions by sometimes asking other questions and how to help people. *Merci Jean.*

## Elements of bibliography

### *Some primary sources on the periodic system*

### *Historical outlines*

***Periodic table in Flatland***

***Periodic table for molecules***

***Quantum-mechanical approaches to the periodic table***

***SO(4,2) approaches to the periodic table***